\documentclass[12pt,preprint]{aastex}

\shorttitle{{\it Chandra} Observation of MS 0839.9+2938}
\shortauthors{Wang et al.} 

\begin{document}

\title{{\it Chandra} Observation of the Cluster of Galaxies MS 0839.9+2938 at $z=0.194$: 
the Central Excess Iron and SN Ia Enrichment}

\author{Yu Wang\altaffilmark{1,2,3}, Haiguang Xu\altaffilmark{1}, 
Zhongli Zhang\altaffilmark{1}, Yueheng Xu\altaffilmark{1}, 
Xiang-Ping Wu\altaffilmark{2}, Sui-Jian Xue\altaffilmark{2},  
and Zongwei Li\altaffilmark{3}}

\altaffiltext{1}{Department of Physics, Shanghai Jiao Tong University, 
1954 Huashan Road, Shanghai 200030, PRC; E-mail: hgxu@sjtu.edu.cn, zebrafish@sjtu.edu.cn}
\altaffiltext{2}{National Astronomical Observatories, Chinese Academy of Sciences, 
20A Datun Road, Beijing 100012, PRC; E-mail: wy@bao.ac.cn, wxp@bao.ac.cn, xue@bao.ac.cn}
\altaffiltext{3}{Department of Astronomy, Beijing Normal University, 
19 Xinjiekouwai Street, Beijing 100875, PRC}

\begin{abstract}
We present the {\it Chandra} study of the intermediately distant ($z=0.194$) cluster 
of galaxies MS 0839.9+2938. By performing both the projected and deprojected spectral 
analyses, we find that the gas temperature is approximately constant at about 4 keV 
in $130-444h_{70}^{-1}$ kpc. In the inner regions, the gas temperature descends towards 
the center, reaching $^{<}_{\sim} 3$ keV in the central $37h_{70}^{-1}$ kpc. This infers 
that the lower and upper limits of the mass deposit rate are  
$9-34 M_{\odot}$ yr$^{-1}$ 
and
$96-126 M_{\odot}$ yr$^{-1}$, respectively 
within $74 h_{70}^{-1}$ kpc
where the gas is significantly colder. Along with the temperature drop, we detect a significant 
inward iron abundance increase from about 0.4 solar in the outer regions to ${\simeq} 1$ solar 
within the central $37h_{70}^{-1}$ kpc. Thus MS 0839.9+2938 is the cluster showing the most 
significant central iron excess at $z ^{>}_{\sim} 0.2$. We argue that most of the excess iron 
should have been contributed by SNe Ia. By utilizing the observed SN Ia rate and stellar mass 
loss rate, we estimate that the time needed to enrich the central region with excess iron is 
$6.4-7.9$ Gyr, which is similar to those found for the nearby clusters. Coinciding with the 
optical extension of the cD galaxy (up to about $30 h_{70}^{-1}$ kpc), the observed 
X-ray surface brightness profile exhibits an excess beyond the distribution expected by 
either the $\beta$ model or the NFW model, and can be well fitted with an empirical two-$\beta$ 
model that leads to a relatively flatter mass profile in the innermost region.
\end{abstract}

\keywords{galaxies: clusters: individual (MS 0839.9+2938) --- intergalactic medium --- galaxies: evolution --- X-rays: galaxies: clusters}

\section{Introduction}
The metallicity of the intra-cluster medium (ICM) in clusters of galaxies is a sensitive 
tracer of both the history of the hierarchical clustering and the chemical evolutions of 
member galaxies and clusters themselves. Since the ICM has been enriched by the supernova 
explosions, the study of the metal abundance gradient in clusters, especially in those 
distant ones, may provide us with crucial constraints on the SN rate, stellar mass loss 
rate, and thus star formation rate. Along with the metal enrichment, a tremendous amount 
of the supernova energy has been inputted into the ICM, primarily in the form of gas 
dynamic energy that may be used to heat the ICM. Thus detailed measurements of the metal 
distribution in clusters as a function of redshift may also provide us with valuable hints 
to investigate how the ICM has been heated, a puzzling question that has plagued astronomers 
for a long time since the absence of massive cooling gas has been confirmed in many 
clusters (see, e.g., Makishima et al. 2001 for a review).

In the past five years, the inward iron abundance increase, first discovered with {\it Ginga}, 
{\it ROSAT} and {\it ASCA} in some nearby clusters that usually host a giant cD galaxy 
(e.g., Koyama et al. 1991; Fukazawa et al. 1994; Ezawa et al. 1997; Xu et al. 1997; Ikebe et al. 1999;
Finoguenov et al. 2000), has been revealed in more sources 
(e.g., a sample of 12 nearby rich clusters at $z ^{<}_{\sim} 0.1$ that have cool cores; 
De Grandi et al. 2004) and in much more details by the 
high spatial resolution observations of {\it Chandra} and {\it XMM}, which allows us to measure 
the abundances of iron and other metals with sufficient accuracy to deduce the SN Ia and SN II 
enrichments and their spatial variations. In clusters that exhibit a central iron excess, it 
has been found that the average iron abundance is typically $0.4-0.6$ solar within a radius 
of 30--80 kpc, which descends to about $0.2-0.3$ solar in the outer regions. In several most 
neighboring sources, such as 
M87/the Virgo cluster (Matsushita et al. 2003), 
NGC 1399/the Fornax cluster (Buote 2002), 
NGC 5044/the NGC 5044 group (Buote et al. 2003), 
NGC 1129/AWM 7 (Furusho et al. 2003), and
NGC 4696/the Centaurus cluster (Sanders \& Fabian 2002), 
the average iron abundance is found to increase further to 0.7--1.5 solar within the central 
15--35 kpc, although in the innermost 2 kpc of M87 and NGC 4696 the iron abundance decreases 
down to $^{<}_{\sim} 0.5$ solar to form a central dip. In M87 (Matsushita et al. 2003), 
NGC 1399 (Buote 2002), NGC 5044 (Buote et al. 2003) and a sample of 19 clusters (Tamura et al. 2004), 
the spatial distributions of the silicon and sulphur abundances are likely to follow that of iron. 
The distribution of the oxygen abundance, on the other hand, is roughly uniform within the errors. 
These results strongly support the idea that there has been an excess SN Ia enrichment in the 
cluster's inner region.

For the time being, the spatial distributions of metal abundances in clusters at $z>0.1$ are 
less understood due to the instrumental limitations, although there is evidence that the same 
central iron abundance excess may also exist (e.g., Allen et al. 2001, 2002; Iwasawa et al. 2001). 
As a step to approach more distant clusters, in this paper we present the {\it Chandra} study 
of the iron abundance distribution in MS 0839.9+2938, an intermediately distant cluster located 
at $z=0.194$ (Stocke et al. 1991). The cluster was listed as an extended X-ray source and 
identified as a cluster in the {\it Einstein} Extended Medium Sensitivity Survey (Gioia et al. 1990). 
It hosts a giant cD galaxy at its dynamic center, which is a weak, compact radio source with 
an intense, extended $H_{\alpha}$ emission (Nesci et al. 1995). The advent of studying this
source is that since its average gas temperature is low ($\simeq 3.6$ keV; see \S4.2), it is
easier to measure the abundance gradient accurately due to the large equivalent widths of 
Fe-$L$ and -$K$ lines.

We organize the paper as follows. In \S2, we describe the observation and data reduction. In 
\S3 and \S4, we present the imaging and spectral analyses, respectively. In \S5, we investigate 
the spatial distributions of the gas and gravitating mass as well as gas cooling. Finally, 
we discuss and summarize our results in \S6 and \S7, respectively. Throughout the paper, we 
adopt the cosmological parameters 
$H_{0} = 70$ km s$^{-1}$ Mpc$^{-1}$, 
$\Omega_{m} = 0.3$ 
and 
$\Omega_{\Lambda} = 0.7$, 
so that $1^{\prime\prime}$ corresponds to $\simeq 3.2h_{70}^{-1}$ kpc at the redshift of the 
cluster. In order to compare directly with previous results, we use the older solar abundances 
from Anders \& Grevesse (1989), where the iron abundance relative to hydrogen is 
$4.68\times 10^{-5}$ 
in number, which is about 46\% higher than the currently preferred meteoritic value 
(McWilliam 1997; Grevesse \& Sauval 1999). Unless stated otherwise, the quoted errors are the 
90\% confidence limits.

\section{Observation and Data Reduction}
The {\it Chandra} observation of MS 0839.9+2938 was carried out on January 29, 2001 for a total 
exposure of 29.7 ks with the chips 2, 3, 6, 7, 8, and 9 of the Advanced CCD Imaging Spectrometer 
(ACIS). The center of the cluster was positioned close to the nominal aim point on the S3 chip 
(CCD 7), so that most of the X-ray emission of the galaxy was covered by the S3 chip. The events 
were collected with frame times of 3.2 s and telemetered in the Faint mode. The focal plane 
temperature was set to be -120 $^{\circ}$C. In this work, we used the {\it Chandra} data analysis 
package CIAO software (version 3.0) to process the data extracted from the S3 chip only. We kept 
events with {\it ASCA} grades 0, 2, 3, 4, and 6, and removed all the bad pixels, bad columns, 
and columns adjacent to bad columns and node boundaries. In order to detect occasional background 
flares, whose effects are particularly significant on the backside-illuminated S1 and S3 chips, 
we examined the lightcurves of the background regions on the S3 chip both in 2.5--7 keV, where the 
background flares are expected most significant, and in 4--7 keV, where the source contamination 
is least. We found that there was almost no strong background flares that increased the count 
rate to over 20\% more than its mean value. By excluding the high background intervals we 
obtained a net exposure of 27.1 ks for the analyses. We restrict our spectral study in 0.7--8.0 keV 
to avoid the calibration uncertainties at the low energy end and the instrumental background at 
the high energy end. In the spectral fittings, we have taken into account corrections for the 
ACIS quantum efficiency degradation and used the XSPEC v11.2.0 software.

\section{Imaging Analysis}
\subsection{X-ray Morphology}
In Figure 1, we present both the smoothed 0.7--8.0 keV image of the central $4.9^{\prime}$ 
($944h_{70}^{-1}$ kpc) of MS 0839.9+2938 in the logarithmic scale, and the corresponding DSS 
optical image on which the X-ray contours are overlaid. The smoothed X-ray image was created 
by using the CIAO tool csmooth with a minimum significance of 3 and a maximum significance 
of 5, and has been background-subtracted and exposure-corrected. As can be seen clearly, the 
diffuse X-ray emission is strongly peaked at the cluster's center. Within the innermost 
$8^{\prime\prime}$ ($\simeq 26h_{70}^{-1}$ kpc), 
it appears to be elongated approximately in the northwest-southeast direction, where it follows 
the distribution of the optical lights. Outside this region the diffuse emission is elongated 
in the north-south direction out to 
$1.1^{\prime}$ ($\simeq 210h_{70}^{-1}$ kpc). 
The X-ray emission of the cluster extends to at least $4.2^{\prime}$ ($\simeq 800h_{70}^{-1}$ kpc),
where the background begins to dominate the flux. There is no apparent substructures in X-rays, 
which indicates that the cluster has not experienced any major merger events in the recent past, 
and is likely to be in a relaxed state. Since the astrometric errors of the {\it Chandra} 
observation are about $1^{\prime\prime}$, the position of the X-ray peak 
(RA=08h42m56.0s, Dec=+29d27m27.2s, J2000) 
is in excellent agreement with the optical center of the cD galaxy 2MASX J08425596+2927272 
(Becker et al. 1995) to within $<0.5^{\prime\prime}$.

Based on the studies of Brandt et al. (2000) and Mushotzky et al. (2000), we estimate that 
there may be 5--10 unrelated X-ray sources in the $r<4.2^{\prime}$ region of the cluster. By 
utilizing a wavedetect detection algorithm and a source detection threshold of 10$^{-6}$, 
and then crosschecking the results both by using the celldetect detection algorithm and by 
eye, we detected 10 X-ray point sources in $r<4.2^{\prime}$, among which less than 1 is 
expected to be a false detection. We find that two of them have optical and radio counterparts 
within $1^{\prime\prime}$. One is CGCG 150-019, an Sb galaxy located at $z=0.0276$ 
(Wegner et al. 2001), and the other is FIRST J084251.9+292825, a Seyfert galaxy in 
MS 0839.9+2938 (Hutchings \& Edwards 2000). In the analysis that follows, all the detected 
X-ray point sources were excluded.

\subsection{Central Excess Emission}
In Figure 2, we show the azimuthally-averaged surface brightness profile of the cluster in
0.7--8 keV that has been corrected with a weighted exposure map. All the detected point sources 
have been excluded, while the background has not been subtracted. We find that in 
$15.8-800h_{70}^{-1}$ kpc, 
the observed X-ray surface brightness profile can be well described ($\chi^2_{\nu} = 1.05$) by 
a model consisting of a standard $\beta$ component (Jones \& Forman 1984) and a spatially uniform 
background as 
\begin{equation}
S(r) =S_{0}[1+(r/r_{\rm c})^{2}]^{0.5-3\beta}+S_{\rm bkg},\\
\end{equation}
where $r_{\rm c}$ is the core radius, $\beta$ is the slope, $S_{\rm bkg}$ is the background, 
and $S_{0}$ is the normalization. The best-fitting parameters are 
$r_{\rm c} = 51.7 \pm 0.5$ $h_{70}^{-1}$ kpc 
and  
$\beta = 0.62 \pm 0.01$. 
However, when we extrapolate the model inwards, the fit becomes worse 
($r_{\rm c} = 41.7\pm0.4$ $h_{70}^{-1}$ kpc,
$\beta = 0.59\pm 0.01$,
and
$\chi^2_{\nu} = 2.84$), 
which significantly underestimates the data in 
$r<4.9^{\prime\prime}$ ($15.8h_{70}^{-1}$ kpc), 
leaving an obvious central excess emission above the model. Such central excess emission has been 
seen in many clusters, e.g., the Centaurus cluster (Ikebe et al. 1999), Abell 1795 (Xu et al. 1998), 
Abell 1983 (Pratt \& Arnaud 2003), and Cl0024+17 (Ota et al. 2004). It may be attributed to either 
a temperature drop, or a metal abundance increase, or an excess of the gas density, all within the 
central region. In \S5.1, we will present detailed model fittings to describe the observed 
X-ray surface brightness profile by taking into account both the temperature and abundance gradients 
obtained from the spectral analyses (\S4).

\section{Spectral Analysis}
\subsection{Two-Dimensional Hardness Ratio Distribution}
Before the direct spectral fittings, we first study the two-dimensional hardness ratio 
distribution of the hot diffuse gas in the cluster. The hardness ratio HR is defined 
as the ratio of the background-subtracted and exposure-corrected counts in 1.4--8.0 keV 
to those in 0.7--1.4 keV. As can be seen in Figure 3, in $\simeq 100-300h_{70}^{-1}$ kpc 
the hardness ratio is approximately constant at a value of $1.00-1.30$, with typical 
errors of $\pm 0.06$. Assuming an average metallicity of 0.4 solar and the Galactic 
absorption 
$N_{\rm H} = 4.04 \times 10^{20}$ cm$^{-2}$ (Dickey \& Lockman 1990),
this is consistent with the hardness ratio of an isothermal plasma at 
$kT \simeq 4.0-6.5$ keV.
In $r \simeq 70-140h_{70}^{-1}$ kpc, there are two relatively colder (90\% confidence level) 
regions located at 
the southeast and northwest of the center, respectively. In the southeast colder region the 
gas temperature is estimated to be $2.8-3.2$ keV (HR$=0.80 \pm 0.04$), which is close to
that of the gas in the northwest colder region ($3.0-3.2$ keV, HR$=0.83 \pm 0.04$).
In $r^{<}_{\sim} 40 h_{70}^{-1}$ kpc, where the hardness ratio is 
$\simeq 0.60-0.83$ with typical errors of $\pm 0.04$, 
the gas appears to be significantly colder than in the outer regions, indicating the 
existence of a plasma at $kT \simeq 2.2-3.1$ keV. These results agree very well with those 
obtained in the detailed spectral fittings (\S4.3).

\subsection{Total Spectrum}
We extracted the total spectrum of the cluster using a circular region filter with a radius 
of $\simeq 2.3^{\prime}$ ($\simeq 440h_{70}^{-1}$ kpc), which is centered at the X-ray peak, 
excluding all the detected point sources. The background spectrum was extracted from a separate 
region on the S3 chip that is far away enough from the cluster. We also have used the $\it Chandra$ 
blank fields for background, and obtained essentially the same results as those shown below.
At the boundary of the so defined cluster region, 
the count flux is about 3 times more than that of the background. The resulting background-subtracted 
spectrum contains a total of about 14,100 counts in 0.7--8.0 keV, showing a strong line 
feature at $\simeq 5.6$ keV that corresponds to a rest-frame energy of about 6.7 keV where 
the K$_{\alpha}$ lines of He-like iron should reside. The appearance of this Fe K$_{\alpha}$ 
feature indicates that the average gas temperature of the cluster should be within 2.7--5.2 keV.

We have fit the spectrum with a model consisting of a single absorbed isothermal component (Table 1). 
First we choose the APEC model (Smith et al. 2001) as the isothermal component, and fix the redshift 
and absorption to $z=0.194$ and the Galactic value, respectively. The fit is marginally acceptable 
($\chi^2_{\nu} = 1.18$), which gives an average gas temperature of
$kT = 3.65^{+0.19}_{-0.20}$ keV, 
and an average metal abundance of  
$Z = 0.44^{+0.09}_{-0.08}$ solar. 
In order to improve the fit we have attempted to apply the VAPEC model instead of the APEC model. 
For MS 0839.9+2938, the redshift is large enough to move the Ly$_{\alpha}$ lines of silicon to lower 
energies to avoid the calibration uncertainties caused by the mirror Ir edge. Thus in the fitting, 
we leave both the Fe and Si abundances free, and tie the abundance of Ni to that of Fe. Although we 
have applied the contamination file from the latest CALDB to correct for the ACIS deficiency 
in low energies, the degradation of energy resolution still does not recover. For an approximate
estimation, we tie the abundances of other $\alpha$-elements (O, Ne, Mg, S, Al, Ar and Ca ) together. 
The model yields a gas temperature of 
$kT = 3.56^{+0.21}_{-0.20}$ keV, 
and abundances of
$Z_{\rm Fe} = 0.46^{+0.10}_{-0.09}$ solar,  
$Z_{\rm Si} = 0.42^{+0.29}_{-0.29}$ solar,
and
$Z_{\alpha} = 1.20^{+0.48}_{-0.42}$ solar
that is mostly dominated by the S abundance.
Although the VAPEC fit ($\chi^2_{\nu} = 1.08$) is better than the APEC fit, there are still 
residuals remaining at about 0.9 keV, where the model underestimates the data. These residuals 
may mostly caused by the excess Fe-$L$ emissions arising from the colder gas in the central 
region, whose existence has been inferred in the study of the hardness ratio distribution (\S4.1). 
In either the APEC or VAPEC fitting, even if the absorption is left free, the obtained column 
density is still consistent with the Galactic value, and the fit is not improved.

We also have examined if the fit to the total spectrum can be further improved by the use of 
a model consisting of two isothermal spectral components, both subjected to a common absorption 
(Table 1). For this purpose we employ two VAPEC components, whose absorption is again fixed to 
the Galactic value because allowing it to vary does not improve the fit. The metal abundances 
of each VAPEC component are grouped in the same way as in the single VAPEC fitting, and the 
abundances of the same element of the two VAPEC components are tied together. We perform the 
F-test and find that at the 95\% confidence level the two-temperature fit ($\chi^2_{\nu} = 1.01$) 
is better than the one-temperature APEC or VAPEC fit. The obtained gas temperatures for the 
two components are 
$kT_{\rm low} = 2.10^{+0.75}_{-1.34}$ keV
and 
$kT_{\rm high} = 6.04^{+4.81}_{-2.24}$ keV, 
and the metal abundances are
$Z_{\rm Fe} = 0.38^{+0.12}_{-0.10}$ solar,  
$Z_{\rm Si} = 0.37^{+0.27}_{-0.24}$ solar,
and
$Z_{\alpha} = 0.96^{+0.47}_{-0.50}$ solar.
With these best-fit parameters we calculate that the unabsorbed 0.5--10 keV flux of the cluster 
is
$3.9_{-0.3}^{+0.2} \times 10^{-12}$ erg cm$^{-2}$ s$^{-1}$,
corresponding to a luminosity of 
$4.1_{-0.2}^{+0.4} \times 10^{44} h_{70}^{-2}$ erg s$^{-1}$.

\subsection{Projected and Deprojected Analyses}
In order to investigate the temperature and abundance gradients, we divided the cluster into 
five annular regions, all centered at the X-ray peak. As the first step, we study the projected 
spectra extracted in these annuli. The background spectrum was again extracted in a region on 
the S3 chip far away enough from the cluster; the use of the {\it Chandra} blank fields yields
essentially the same results. Assuming that the gas in each annulus is in the thermal equilibrium, 
we fit the spectra using a single absorbed APEC model with the absorption fixed to the Galactic 
value; allowing the absorption to vary does not improve the fits. As is shown in Figure 4 and 
Table 2, the fit is acceptable for each region. The gas temperature is found to be approximately 
constant at $\simeq 4.3$ keV outside about $130h_{70}^{-1}$ kpc, while it descends to 
$\simeq 3.2$ keV in the central $37h_{70}^{-1}$ kpc, as has been expected in the study of the 
hardness ratio distribution (\S4.1). On the other hand, the average metal abundance, which 
is mainly determined by iron, increases towards the center from a roughly constant value of 
0.4 solar in the outer regions to about 0.8 solar within the central $37h_{70}^{-1}$ kpc. 
Instead of the APEC model, we have also attempted to apply the VAPEC model (Fig. 4 and Table 2). 
For better statistics, all the abundances of the $\alpha$-elements are tied together, and the 
absorption is again fixed to the Galactic value. We find that the fits are acceptable too, 
and the obtained gas temperatures agree nicely with those obtained with the APEC model. In 
the regions outside $r = 74h_{70}^{-1}$ kpc, the iron abundance varies slowly in a range of 
0.25--0.42 solar, while in the central region, it increases drastically to about 0.8 solar. 
The spatial distribution of the average abundance of the $\alpha$-elements, however, is not 
well determined due to large errors.

Next, we perform the deprojected analysis of the spectra extracted in the inner four regions 
by using the standard "onion peeling" method (Fig. 4 and Table 2). Assuming that the gas 
temperature and abundance are uniform in each spherical shell, and that the gas density follows
the best-fit profile derived in \S5.1, we subtract emissions from the outer shells projected 
on the inner shells. We fit the deprojected spectra with either the APEC or VAPEC 
model whose absorption is fixed to the Galactic value. We find that the deprojected spectra 
can be well fitted with either of the models. The gas temperatures obtained with the APEC 
model agree very well with those obtained with the VAPEC model, and both the average metal 
abundance of the APEC model and the iron abundance of the VAPEC model show significant 
central increase in the innermost region as has been found in the projected analysis. For 
each model the best-fit gas temperatures and metal abundances obtained in the projected 
and deprojected analyses are consistent with each other. In the innermost region, the 
deprojected temperatures are only slightly lower than, but still consistent with the 
projected values.

\subsection{Temperature and Iron Abundance Gradients in the ICM}
As we have shown above, both the projected and deprojected spectral analyses indicate that 
in MS 0839.9+2938 there is a central temperature drop within a radius of about $74h_{70}^{-1}$ kpc. 
According to the deprojected APEC fit, the gas temperature descends from $kT \simeq 4.2$ keV 
in $> 130h_{70}^{-1}$ kpc to $\simeq 2.7$ keV in $< 37h_{70}^{-1}$ kpc, which can be approximated
by a power-law profile 
$T(R)=T_{0} (R/R_{0})^{\alpha}$, 
where $T_{0}=3.0$ keV, $\alpha = 0.17$ and $R_{0} = 35.4h_{70}^{-1}$ kpc (Fig. 4a). In 
the same region where the cold gas is present, the iron abundance shows a clear tendency to 
increase. In the projected analysis, both the iron-dominated metal abundance of the APEC model 
and the iron abundance of the VAPEC model in the central $37h_{70}^{-1}$ kpc region are significantly 
higher than their counterparts in 
$74-130h_{70}^{-1}$ kpc 
at the 90\% confidence level. They are also apparently higher than the abundances in 
$37-74h_{70}^{-1}$ kpc 
and 
$130-230h_{70}^{-1}$ kpc, 
respectively, with the 90\% confidence ranges overlap with each other only slightly. In the 
deprojected analysis, although the obtained errors are much larger, the average metal abundance 
and iron abundance still show a tendency to increase inward. At the 90\% confidence level, their 
values inferred for the 
$< 37h_{70}^{-1}$ kpc 
region are significantly higher than those for the
$74-130h_{70}^{-1}$ kpc 
region.

In order to crosscheck these results with better statistics, we divide the cluster into 
the inner 
($<74h_{70}^{-1}$ kpc) 
and outer
($74-444h_{70}^{-1}$ kpc)
regions, and fit their deprojected spectra with the an absorbed VAPEC model (Table 3). 
In the fitting the absorption is fixed to the Galactic value, and the metal abundances
are grouped in the same way as in the fittings of the total spectrum. We find that the fits 
are acceptable for both regions. At the 90\% confidence level, the central temperature 
decrease and abundance increase can be confirmed. The silicon abundance and the average 
abundance of other $\alpha$-elements, however, show no tendency to vary between the two 
regions.

Since the iron abundance is determined as much by the Fe-L emission lines keV as by 
the Fe-K lines, the obtained inward iron abundance increase might be an artifact, which 
is actually caused by the enhancement of the Fe-L blend due to an inward temperature 
drop that is even steeper than has been modeled. We have examined this possibility. 
First we exclude the Fe-$L$ lines completely from the spectra and use the Fe-$K$ lines 
only to measure the iron abundance. The obtained iron abundance does show a tendency 
to increase inwards, although it is less significant due to poorer statistics. Second, 
we fit the deprojected spectrum of the innermost region by using a single absorbed 
VAPEC model with the absorption fixed to the Galactic value and the iron abundance 
fixed to 0.47 solar (the upper limit of the iron abundance in the third annular region; 
Table 2). We do obtain a lower temperature of $2.38 \pm 0.30$ keV as is expected. 
However, the fit appears to be worse and should be ruled out at the 90\% confidence 
level in terms of F-test, showing obvious spectral residuals in $4.5-7.0$ keV as well as at 
$\simeq 0.9$ keV, where the fits to the Fe-$K$ and Fe-$L$ lines are worse. This indicates 
that a central abundance increase is much more preferred than an even steeper temperature 
gradient. Third, we study the two-dimensional fit-statistic contours of temperature 
and iron abundance at the 68\%, 90\% and 99\% confidence levels for the innermost region 
and the third region, respectively, all obtained in the deprojected VAPEC fitting (Fig. 5). 
As can be clearly seen the central iron abundance is higher than that of the third region at 
a significance of 99\%. Consequently, we conclude that the possibility of the detected central 
iron abundance increase being an artifact should be excluded. Note that the coexistence of 
the temperature drop and abundance increase may indicate that they are physically correlated, 
since the higher the metallicity is, the faster the gas cools down.

\section{Mass Distributions and Cooling of Gas}
\subsection{Gas Density Profile}
Assuming that 1) the gas is ideal and in the thermal equilibrium state, and 2) the spatial 
distribution of the gas is spherically asymmetric, we deduce the three-dimensional gas 
density distribution $n_{g}(R)$ by fitting the observed radial X-ray surface brightness profile 
$S(r)$, which was extracted in 0.7--8.0 keV, azimuthally-averaged, and exposure-corrected. 
The calculations was performed in a projected way by using
\begin{equation}
S(r) = S_{0} \int^{\infty}_{r} \Lambda (T,Z) n_{g}^{2}(R) \frac{RdR}{\sqrt{R^{2}-r^{2}}}+S_{\rm bkg},
\end{equation}
where $\Lambda(T,Z)$ is the cooling function calculated by taking into account both the 
temperature and abundance gradients as is modeled by the best-fit deprojected APEC parameters, 
and $S_{\rm bkg}$ is the background.

First we assume that the gas density follows the $\beta$ distribution
\begin{equation}
n_{g}(R) = n_{g,0} \left[ 1 + (R/R_{c})^{2} \right] ^{-3\beta/2},
\end{equation}
where $R_{c}$ is the core radius and $\beta$ is the slope. We find 
$R_{c} = 47.3 \pm 0.3 h_{70}^{-1}$ kpc
and
$\beta = 0.61 \pm 0.01$. 
The fit, however, is unacceptable ($\chi^2_{\nu} = 1.63$), showing strong residuals in 
$<15.8 h_{70}^{-1}$ kpc 
where the model significantly underestimates the data (Fig. 2). In fact, the gas density 
derived in such a way is lower than that obtained in the deprojected spectral analysis 
(\S4.3) by about 45\%. This confirms the result in \S3.2, although the central brightness 
excess above the model is less prominent since the effects of the central temperature 
drop and abundance increase are both counted in, which enhance the model emission in 
the central region.

Next we examine if $S(r)$ can be reproduced by adopting the universal gravitating mass 
distribution as has been found in the cosmological N-body simulations (Navarro et al. (1997) 
\begin{equation}
\rho (R) = \frac{\delta_{c}\rho_{crit}}{(R/R_{s})(1+R/R_{s})^{2}},
\end{equation}
where $\rho(R)$ is the gravitating mass density, $R_{s}$ is the scale radius, $\rho_{crit}$ 
is the critical density of the universe at the observed redshift, and $\delta_{c}$ is the 
characteristic density contract that can be expressed in term of the equivalent concentration 
parameter {\it c} as
\begin{equation}
\delta_{c} = \frac{200}{3} \frac{c^{3}}{[{\rm ln}(1+c)-c/(1+c)]}.
\end{equation}
As is shown in Makino et al. (1998) and Wu et al. (2000), under the assumptions of 1) spherical 
symmetry, 2) hydrostatic equilibrium, and 3) ideal and isothermal gas,
the gas distribution can be given in an analytic form
\begin{equation}
\tilde{n}_{g}(x) = \tilde{n}_{g}(0) \frac {(1+x)^{\alpha/x}-1}{e^{\alpha}-1},
\end{equation}
where $x = R/R_{s}$ is the dimensionless radius, 
$\alpha = 4\pi G \mu m_{p} \delta_{c} \rho_{crit} R_{s}^{2}/kT$ 
is the index, and $\mu = 0.609$ is the average molecular weight for a fully ionized gas. Note
that in Eq.(6) a background density at infinity has been subtracted to avoid the divergence in 
integrating the X-ray emission along the line of sight. As are shown in Figure 2 
and Table 4. We find that, comparing with the $\beta$ model, the fit is improved in the central 
region, giving
$R_{s} = 195.1 \pm0.5 h_{70}^{-1}$ kpc, 
$c = 6.5 \pm 0.1$, 
and
$\chi^2_{\nu} = 1.12$.
However, it still systematically underestimates the data in $<15.8$ kpc.

In order to improve the fit in the central region, we have attempted to add an additional 
$\beta$ component to the $\beta$ model to account for the central excess emission, for 
which the gas density distribution is
\begin{equation}
n_{g}(R) = \left\{
           n_{g1,0}^{2} \left[ 1 + (R/R_{c1})^{2} \right] ^{-3\beta_{1}}
         + n_{g2,0}^{2} \left[ 1 + (R/R_{c2})^{2} \right] ^{-3\beta_{2}}
           \right\} ^{1/2}.
\end{equation}
This empirical two-component model has been successfully applied to many clusters of galaxies 
as well as some groups of galaxies (see, e.g., Makishima et al. 2001 and references therein). We 
find that the model can best fit the observed surface brightness profile throughout the spatial 
range spanned by the cluster and yield an acceptable fit. The best-fit parameters are 
$R_{c1} = 65.9_{-0.5}^{+0.6} h_{70}^{-1}$ kpc 
and 
$\beta_{1}=0.66^{+0.01}_{-0.01}$
for one $\beta$ component, which mostly describes the surface brightness profile in the outer 
region, and
$R_{c2}=15.5_{-0.4}^{+0.3}$$h_{70}^{-1}$ kpc 
and 
$\beta_{2}=0.59^{+0.01}_{-0.01}$
for another, which describes the central excess brightness.

\subsection{Gas and Gravitating Mass Distributions}
The distribution of the diffuse X-ray emission in different energy bands is roughly 
symmetric outside to about $4.2^{\prime}$ ($\simeq 800h_{70}^{-1}$ kpc) in MS 0839.9+2938, 
indicating that the cluster is nearly relaxed. In this case the X-ray imaging spectroscopy 
can be considered as a reliable tool to deduce the mass distributions in the cluster. 
Assuming that the gas is ideal and the cluster is in the hydrostatic equilibrium state, 
and utilizing the best-fit temperature gradient obtained in the deprojected analyses and 
the gas density profile deduced with the two-$\beta$ model, we calculate the three-dimensional 
gravitating mass distribution by using
\begin{equation}
M(R) = -\frac{kTR}{G\mu m_{p}}\left(\frac{d\rm ln \it n_{g}}{d\rm ln \it R}+\frac{d\rm ln \it T}{d\rm ln \it R}\right),
\end{equation}
where $M(R)$ is the total mass within the radius $R$, G is the gravitational constant, and 
$m_{p}$ is the proton mass. Here the effects of non-thermal turbulent, magnetic 
pressure, and cosmic ray pressure are not taken into account. The resulting distribution of 
the total gravitating mass is shown in Figure 6, along with the 90\% errors determined by 
performing the Monte-Carlo simulations that account for the range of temperature and gas profiles 
allowed by the data. The total mass distribution shows an excess beyond that predicted by the 
$\beta$ model in $R^{<}_{\sim} 30 h_{70}^{-1}$ kpc, which is roughly the optical boundary of 
the cD galaxy (Nesci et al. 1989) and the region where the excess X-ray emission, temperature 
drop, and iron abundance increase are observed. The excess mass  
($\simeq 6 \times 10^{11} M_{\odot}$)
is comparable to the mass of the cD galaxy as is inferred from its optical luminosity 
($\simeq 1 \times 10^{12} M_{\odot}$ within $30 h_{70}^{-1}$ kpc; Nesci et al. 1989). At the 
virial radius 
$R_{200}$ ($1640h_{70}^{-1}$ kpc), 
within which the average mass density is 200 times the current critical density of the universe, 
the extrapolated gravitating mass is 
$M_{200} = 6.1 \times 10^{14} M_{\odot}$. 
In Figure 6, we also show the calculated gas mass fraction along with the 90\% errors. It 
increases from about 0.05 at the center to about 0.1 at $\simeq 100h_{70}^{-1}$ kpc, and then 
keeps as a constant in the outer regions. For $\Omega_{b}h^{2} = 0.024\pm0.001$ (Spergel et al. 2003), 
we obtain $\Omega_{m} = 0.42_{-0.14}^{+0.13}$, which is self-consistent with the value adopted
in this work.

\subsection{Cooling of Gas and Mass Deposit Rate}
With the gas density $n_{g}(R)$ deduced with the best-fit two-$\beta$ model (\S5.1), we 
estimate the gas cooling time as
\begin{equation}
t_{\rm cool} = 2.4 \times 10^{10}\rm yr \left( \frac{\it kT}{\rm keV} \right)^{1/2} \left( \frac{ {\it n_{g}}}{10^{-3} \rm cm^{-3}} \right) ^{-1}
\end{equation}
(cf, Sarazin 1986), where both continuum and line emissions are included (Fig. 6). We find 
that 
$t_{\rm cool} \sim 10^{8}$ yr 
at the center, which is much shorter than the age of the universe at the cluster's redshift. 
The cooling radius $R_{\rm cool}$, defined as the radius where the cooling time is equal to 
the age of the universe, is $\simeq 150 h_{70}^{-1}$ kpc. This is roughly the 
radius within which the gas temperature begins to descend inwards.

In order to determine the mass deposit rate of the cooling processes, we fit the spectrum 
extracted in the two innermost annuli ($<74.1 h_{70}^{-1}$ kpc) where the gas temperature
is significantly lower than that of the outer regions. We use the cooling flow model 
MKCFLOW, which is absorbed by a column density fixed to the Galactic value (Fig. 7). We find 
that when the low temperature $kT_{\rm low}$ of MKCFLOW is fixed to 0.08 keV, the model yields 
a rather poor fit to the data ($\chi^2_{\nu} = 1.69$), giving rise to too much cool gas and 
thus overestimating the data in $^{<}_{\sim} 0.8$ keV. When $kT_{\rm low}$ is left free, 
the fit becomes acceptable ($\chi^2_{\nu} = 1.00$), but it requires 
$kT_{\rm low}=1.14^{+0.31}_{-0.42}$ keV,
which may suggest that the spectrum is dominated by a hot component. We then add an additional 
ambient isothermal component (MEKAL) into the model, tying its temperature to $kT_{\rm high}$ 
of MKCFLOW. When 
$kT_{\rm low}$ is fixed to 0.08 keV, we obtain  
$kT_{\rm high}=3.40^{+0.27}_{-0.25}$ keV and
$\dot{M} = 53^{+43}_{-44} M_{\odot}$ yr$^{-1}$; 
when $kT_{\rm low}$ is free, we obtain 
$kT_{\rm low}=0.56^{+1.53}_{-0.48}$ keV,
$kT_{\rm high}=3.42^{+0.25}_{-0.19}$ keV, 
and
$\dot{M} = 67^{+59}_{-33} M_{\odot}$ yr$^{-1}$. 
Whether or not $kT_{\rm low}$ is fixed, the MKCFLOW+MEKAL fit is acceptable 
($\chi^2_{\nu} = 0.99$ and $1.00$, respectively), but $kT_{\rm low}$ and $\dot{M}$ are 
far from being well constrained. On the other hand, the two-temperature APEC model  
also gives a good fit ($\chi^2_{\nu} = 0.99$) to the data. The derived best-fit low 
and high temperatures are 
$kT_{\rm low} = 2.17^{+1.08}_{-0.66}$ keV and 
$kT_{\rm high} = 4.85^{+0.65}_{-1.38}$ keV. 
Thus, given the current data quality and instrumental capabilities, the two-phase isothermal 
model cannot be distinguished from the single MKCFLOW model with a free $kT_{\rm low}$, or the 
MEKAL+MKCFLOW model. Although not well constrained, the cooling flow model cannot be ruled 
out for this cluster.

\section{Discussions}
\subsection{Metal Enrichment in ICM and Role of SNe Ia in Inner Region}
The inward iron abundance increase found in many nearby clusters and groups around/in 
the giant, dominating elliptical galaxy at the center seems to follow a somewhat universal 
profile. Actually, when we compare the iron abundances in the inner ($^{<}_{\sim} 70h_{70}^{-1}$) 
and outer regions ($Z_{\rm Fe,in}$ and $Z_{\rm Fe,out}$, respectively) of the clusters 
in which an inner abundance enhancement is observed (Buote et al. 2003; Nevalainen et al. 2001; 
Pratt $\&$ Arnaud, 2003; Takahashi $\&$ Yamashita 2003; Finoguenov et al. 2004; Tamura et al. 2004), 
we find that within the errors the ratio $Z_{\rm Fe,in}/Z_{\rm Fe,out}$ is roughly a constant 
as a function of either the ICM temperature or the cluster's B-band luminosity (Fig. 8). 
All these results strongly imply that the observed central iron excess has probably 
been caused by the same enrichment process associated with the central dominating galaxy, 
which is a cD galaxy in most cases. Meanwhile, the high spatial resolution observations 
of the nearby giant ellipticals M87 (Matsushita et al. 2003) and NGC 5044 (Buote et al. 2003) 
revealed that the spatial distribution of oxygen abundance is approximately uniform out to 
about 40--70 kpc, inferring that the metals synthesized in SNe II are evenly distributed 
in the cluster/group. Thus, the central iron abundance increase should be ascribed mainly 
to SNe Ia.

By performing both the projected and deprojected spectral analyses, we find that in MS 0839.9+2938 
the measured average iron abundance is about 0.4 solar in the outer regions, while it increases 
to ${\simeq}1$ solar within the central $37 h^{-1}_{70}$ kpc at the 90\% confidence level. 
Reports about detections of such a significant central iron excess in clusters at $z ^{>}_{\sim} 0.2$ 
are rare in literatures. In 4C+55.16 ($z=0.240$; Iwasawa et al. 2001) the metal abundance increases 
from $^{<}_{\sim} 0.5$ solar in outer regions to $>1$ solar within the central $34 h_{70}^{-1}$ kpc. 
In Abell 2390 ($z=0.23$; Allen et al. 2001), it increases from $0.23^{+0.12}_{-0.13}$ 
solar in $143-714 h_{70}^{-1}$ kpc to $0.48^{+0.11}_{-0.10}$ solar within $<71 h_{70}^{-1}$ kpc. 
However, note that the results of the latter two cases were acquired by performing projected analysis 
at the 68\% confidence level, and the abundance is an average value for both iron and $\alpha$-elements, 
although it is iron-dominated.

In MS 0839.9+2938, the excess iron mass within $R<74h_{70}^{-1}$ kpc is estimated to be 
$M_{\rm Fe,excess} \simeq 3.0 \times 10^{8} M_{\odot}$,
which is similar to that of Abell 2390 
($M_{\rm Fe,excess} \simeq 2.5 \times 10^{8} M_{\odot}$ 
within $R<71h_{70}^{-1}$ kpc, based on the data of Allen et al. 2001). The derived $M_{\rm Fe,excess}$ 
obviously exceeds the iron enrichment caused only by the stellar winds and ram pressure stripping 
of metal-enriched gas (e.g., Trager et al. 2000; Ciotti et al. 1991). Therefore, we speculate that 
a significant part of the excess iron may have been blown into the ICM directly by the SN Ia explosions.

By reproducing the observed iron-to-silicon abundance ratio in the central 
$74 h^{-1}_{70}$ kpc of MS 0839.9+2938 
(Fe/Si=$1.29_{-0.52}^{+1.40}$ solar at the 68\% confidence level), where a clear iron abundance 
increase is seen, we have attempted to estimate the SN Ia contribution to the iron enrichment. 
The current supernova models give very different yields to the metals (Gibson et al. 1997). 
When we adopt the theoretical SN Ia yields obtained with the W7 model and the weighted SN II yields 
in Nomoto et al. (1997), where an IMF slope of $x = 1.35$ (Salpeter 1955) and a stellar mass range 
of $10-50 M_{\odot}$ were assumed for the SN II progenitors, we find that the SN Ia contribution 
to the total iron mass is $90^{+10}_{-18}\%$, corresponding to an iron abundance of 
$Z_{\rm Fe,SN Ia}=0.56^{+0.06}_{-0.11}$ solar. 
This fraction is consistent with those found in NGC 5044 ($67-79$\%; Buote et al. 2003), M87 
($55-97$\%; Matsushita et al. 2003), 
NGC 1399 ($\simeq 80\%$; Buote 2002), and the RGH 80 group ($\simeq 80\%$; Xue et al. 2004). We 
are not able to derive reasonable results by using the SN Ia yields obtained with either the WDD1 
or WDD2 model in Nomoto et al. (1997), because both models give rise to too much silicon.

Using the same method as above we estimate that for the $74 < R < 444h_{70}^{-1}$ kpc region,
where the observed Fe/Si ratio is $0.84_{-0.47}^{+2.71}$ solar (68\% confidence), the SN Ia 
contributions is $75^{+25}_{-55}\%$, or $Z_{\rm Fe,SN Ia}=0.23^{+0.08}_{-0.17}$ (W7 model). 
By comparing these results with those for the inner region, we find that the spatial distribution 
of the SN II contribution to iron is roughly constant at about $0.08$ solar, which is close 
to that in Finoguenov et al. (2002; $Z_{\rm Fe,SN II}=0.1-0.15$ solar) and Matsushita et al. 
(2003; $Z_{\rm Fe,SN II}=0.2-0.4$ solar).

It is crucial to examine if there has been enough time for SNe Ia to enrich the cluster's 
central region with the observed amount of excess iron. Since the SN Ia enrichment of iron 
in the ICM is determined both by iron blown out directly into the ICM during the SN Ia explosions 
and by iron lost in the stellar winds, we calculate the enrichment time $t_{\rm enrich}$ 
(B$\ddot{\rm o}$hringer et al. 2004) using 
\begin{equation}
t_{\rm enrich} = M_{\rm Fe, SN Ia} / [L_{\rm B} \times (R_{\rm SN Ia}+R_{\rm wind})], 
\end{equation}
where $M_{\rm Fe, SN Ia}$ is the iron mass contributed by SNe Ia,
$L_{\rm B}$ ($2.2 \times 10^{11} h_{70}^{-2} {\rm L_{\odot}}$; Nesci et al. 1989) 
is cluster's B-band luminosity, $R_{\rm SN Ia}$ is the direct SN Ia iron-enriching rate in units of 
$M_{\odot}$ yr$^{-1}$ $L_{\odot}^{-1}$,
and 
$R_{\rm wind}$ is the iron-enriching rate due to the stellar mass loss in units of 
$M_{\odot}$ yr$^{-1}$ $L_{\odot}^{-1}$.
As is shown above, we assume that SNe II evenly enriched the gas throughout cluster, 
yielding an iron abundance of 0.08 solar, so by subtracting the SN II contribution, we obtain  
$M_{\rm Fe, SN Ia} = 5.2 \pm 0.8 \times 10^{8} M_{\odot}$.
In SN Ia explosions the rate of iron mass that has been directly transferred into the ICM is 
\begin{equation}
R_{\rm SN Ia} = SR 10^{-12}{\rm L^{-1}_{B,\odot}} \eta _{\rm Fe},
\end{equation}
where $\eta_{\rm Fe}$ is the iron yield per SN Ia event, and $SR$ is the SN Ia rate in units of 
SNu 
(1 SNu = 1 supernova $(10^{10} L_{B,\odot})^{-1}$ century$^{-1}$). 
In calculations, we choose $\eta_{\rm Fe} = 0.5$ and 0.7 $M_{\odot}$, respectively, 
and adopt either a temporally constant $SR$ of $0.18h^{2}_{70}$ SNu (Cappellaro et al. 1999), 
or an observation-constrained time-dependent $SR$ that increases with the redshift until 
$z \simeq 1$ and then drops (Dahlen et al. 2004). In the stellar winds, the iron loss rate is
\begin{equation}
R_{\rm wind} = 1.5 \times 10^{-11} {\rm L^{-1}_{B,\odot}}t_{15}^{-1.3} \gamma_{\rm Fe},    
\end{equation}
where $\gamma _{\rm Fe}$ is the iron mass fraction in stellar winds, and $t_{15}$ is the age 
in units of 15 Gyr (Ciotti et al. 1991). We list the calculated enrichment times and redshifts 
at which the SN Ia enrichment is expected to start in Table 5. We find that when the supernova 
rate was a constant in the past, the derived enrichment time is too close to or even longer than 
the Hubble time at the cluster's redshift ($11.1 \times 10^{9}$ yr). When the observed time-dependent 
supernova rate provided in Dahlen et al. (2004) is adopted, the derived enrichment times 
are 7.9 Gyr and 6.4 Gyr for 
$\eta=0.5$ and $0.7$ $M_{\odot}$, respectively. For comparison, we also perform similar calculations 
for the inner $71h_{70}^{-1}$ kpc region of Abell 2390 and find that the resulting enrichment time 
is 6.6--10.7 Gyr. These results agree with those deduced for the Virgo cluster, the Centaurus cluster, 
the Perseus cluster and Abell 1795 
($t_{\rm enrich} = 5-11$ Gyr; B$\ddot{\rm o}$hringer et al. 2004), 
indicating that the excess iron can be entirely produced by the central brightest galaxy 
(De Grandi \& Molendi 2001; De Grandi et al. 2004). Since only few SNe Ia are observed at 
$z ^{>}_{\sim} 1.8$, we predict that the most distant clusters showing a significant central 
iron excess should be detected at no farther than $z \simeq 0.5$.

\subsection{Scaling Relations}
The derived X-ray temperature, X-ray luminosity and virial mass at $R_{200}$ of MS 0839.9+2938 
are 
$kT = 3.65_{-0.20}^{+0.19}$ keV,
$L_{\rm X} = 4.1_{-0.2}^{+0.4} \times 10^{44}$ $h_{70}^{-2}$ erg s$^{-1}$,
and
$M_{200} = 6.1\times 10^{14} h_{70}^{-1} M_{\odot}$, 
respectively. Within the errors, these values are consistent with the observed luminosity-mass 
relation (e.g., Markevitch et al. 1998) and mass-temperature relation (e.g., Sanderson et al. 2003) 
after the effect of gas cooling is compensated. Using the gas density obtained with the best-fit 
two-$\beta$ model (\S5.1) and the analytical power-law temperature profile (\S4.4), we calculate 
the azimuthally averaged gas entropy $S=kT/n^{2/3}$ in the cluster and plot its spatial 
distribution in Figure 9. In $R>0.03 R_{200}$ ($49h_{70}^{-1}$ kpc), the profile can be approximated 
by a power-law form with an index of 1.1, as has been expected from shock heating that occurred 
during the spherical collapse (Tozzi \& Norman 2001). At $0.1R_{200}$, the derived gas entropy is 
$\simeq 167_{-32}^{+22} h_{70}^{-1/3}$ keV cm$^{2}$, 
which is consistent with the entropy-temperature relation for a sample of 66 relaxed clusters 
presented in Ponman et al. (2003).

\section{Summaries}
We present the {\it Chandra} ACIS study of the intermediately distant ($z=0.194$) cluster of 
galaxies MS 0839.9+2938. By performing both the projected and deprojected spectral analyses, 
we find that the emission-measure weighted gas temperature is approximately constant at about 
4 keV in $>130h_{70}^{-1}$ kpc. In the inner regions, the gas temperature descends towards the 
center, reaching $^{<}_{\sim} 3$ keV within the innermost $37h_{70}^{-1}$ kpc. Along with the 
temperature drop, we detect a significant inward iron abundance increase from about 0.4 solar 
in the outer regions to ${\simeq} 1$ solar within the central $37h_{70}^{-1}$ kpc. Besides 
MS 0839.9+2938, another cluster at $z \simeq 0.2$ that has been found to show a similar, strong 
central iron excess is Abell 2390 ($z=0.23$; Allen et al. 2001). We argue that most of the 
excess iron is contributed by SNe Ia. By using the observed SN Ia rate and stellar mass loss 
rate, we estimate that in MS 0839.9+2938 the time needed to enrich the central region with 
excess iron is $6.4-7.9$ Gyr, which is similar to those found for the nearby clusters. In 
almost the same region where the temperature drop and abundance increase are seen, the observed 
X-ray surface brightness profile shows an excess beyond the distribution expected by either the 
$\beta$ model or the NFW model, and can be well fitted with an empirical two-$\beta$ model. 
The origins of all these phenomena can be correlated with each other (cf, Makishima et al. 2001). 
The excess amount of gas, a part of which comes from the supernova explosions with heavy metals, 
is trapped around the central dominating galaxy to form the observed surface brightness excess. 
The relatively high gas density and high metallicity both speed up the gas cooling process, 
causing the observed inward gas temperature drop.

\acknowledgments 
This work was supported by the National Science Foundation of China (Grant No. 10273009 \& 10233040), 
the Ministry of Science and Technology of China (Grant No. NKBRSF G19990754), and Shanghai Key 
Projects in Basic Research (No. 04JC14079).


\begin{deluxetable}{lccccccr}
\tablecaption{Fits to the Total Spectrum} \label{tbl:totalspectrum}
\tabletypesize{\scriptsize}
\tablewidth{0pt}
\tablehead{
\colhead{Model} & 
\colhead{$N_{\rm{H}}$} & 
\colhead{$kT$} &
\colhead{$Z$} & 
\colhead{$Z_{\alpha}$ \tablenotemark{a} \,} &
\colhead{$Z_{\rm{Si}}$} & 
\colhead{$Z_{\rm{Fe}}$} &
\colhead{$\chi^2_{\nu}$} \\ 
& 
\colhead{$(10^{20}$ cm$^{-2})$} & 
\colhead{(keV)} & 
\colhead{(solar)} & 
\colhead{(solar)} &
\colhead{(solar)} & 
\colhead{(solar)}
}
\startdata
APEC &4.04 (Fixed)          &$3.65^{+0.19}_{-0.20}$&$0.44^{+0.09}_{-0.08}$&$-$&$-$&$-$&107.6/91\\
     &$1.95^{+1.57}_{-1.53}$&$3.88^{+0.25}_{-0.26}$&$0.46^{+0.10}_{-0.09}$&$-$&$-$&$-$&102.8/90\\
\hline
VAPEC&4.04 (Fixed)          &$3.56^{+0.21}_{-0.20}$&$-$                   &$1.20^{+0.48}_{-0.42}$&$0.42^{+0.29}_{-0.29}$&$0.46^{+0.10}_{-0.09}$&96.4/89\\ 
     &$1.55^{+1.67}_{-1.55}$&$3.83^{+0.26}_{-0.28}$&$-$                   &$1.42^{+0.61}_{-0.50}$&$0.62^{+0.41}_{-0.35}$&$0.50^{+0.11}_{-0.11}$&90.5/88\\
\hline
2$-$VAPEC&4.04 (Fixed)&\multicolumn{2}{l}{$6.04^{+4.81}_{-2.24}$(High) $-$}&$0.96^{+0.47}_{-0.50}$&$0.37^{+0.27}_{-0.24}$&$0.38^{+0.12}_{-0.10}$&87.7/87\\
         &            &\multicolumn{2}{l}{$2.10^{+0.75}_{-1.34}$(Low)}&&&&\\
\enddata
\tablenotetext{a}{Average abundance of the $\alpha$-elements except for Si.} 
\end{deluxetable}

\begin{deluxetable}{llccccccccc}
\tablecaption{Projected and Deprojected Spectral Analyses \tablenotemark{a} \,\label{tbl:spectra}}
\tabletypesize{\scriptsize}
\tablewidth{0pt}
\tablehead{
\multicolumn{11}{c}{APEC Model} \\ \cline{1-11}
\colhead{Region} &
\colhead{Radius} &
\multicolumn{4}{c}{Projected} & &
\multicolumn{4}{c}{Deprojected}\\ 
\cline{3-6}
\cline{8-11}
&  &
\colhead{$kT$} &
\multicolumn{2}{c}{$Z$} & 
\colhead{$\chi^2_{\nu}$} & &
\colhead{$kT$} & 
\multicolumn{2}{c}{$Z$} &
\colhead{$\chi^2_{\nu}$} \\
&
\colhead{($h_{70}^{-1}$ kpc)} &
\colhead{(keV)} &
\multicolumn{2}{c}{(solar)} &
& &
\colhead{(keV)} &
\multicolumn{2}{c}{(solar)} & 
}
\startdata
1&0.0-37.3   &$3.16^{+0.19}_{-0.20}$&\multicolumn{2}{c}{$0.80^{+0.23}_{-0.20}$}&44.0/43&&$2.69^{+0.49}_{-0.32}$&\multicolumn{2}{c}{$1.15^{+0.95}_{-0.52}$}&34.9/43  \\
2&37.3-74.1  &$3.46^{+0.29}_{-0.25}$&\multicolumn{2}{c}{$0.47^{+0.16}_{-0.15}$}&40.8/44&&$3.15^{+0.43}_{-0.42}$&\multicolumn{2}{c}{$0.60^{+0.38}_{-0.28}$}&34.2/44  \\
3&74.1-130.1 &$3.80^{+0.35}_{-0.35}$&\multicolumn{2}{c}{$0.26^{+0.16}_{-0.14}$}&36.2/43&&$3.49^{+0.63}_{-0.56}$&\multicolumn{2}{c}{$0.20^{+0.26}_{-0.20}$}&32.3/43  \\
4&130.1-230.0&$4.27^{+0.62}_{-0.41}$&\multicolumn{2}{c}{$0.37^{+0.24}_{-0.20}$}&41.0/38&&$4.23^{+0.81}_{-0.57}$&\multicolumn{2}{c}{$0.40^{+0.35}_{-0.27}$}&45.7/43  \\
5&230.0-444.1&$4.29^{+1.10}_{-0.76}$&\multicolumn{2}{c}{$0.42^{+0.52}_{-0.33}$}&35.9/42&&$-$                   &\multicolumn{2}{c}{$-$}                   & $-$     \\
\hline
\multicolumn{11}{c}{VAPEC Model} \\ \cline{1-11}
 & &
\multicolumn{4}{c}{Projected} & &
\multicolumn{4}{c}{Deprojected} \\ 
\cline{3-6}
\cline{8-11}
 & &$kT$ &$Z_{\alpha}$&$Z_{\rm{Fe}}$&$\chi^2_{\nu}$ &                                    &$kT$ &$Z_{\alpha}$&$Z_{\rm{Fe}}$& $\chi^2_{\nu}$  \\
 & &(keV)&(solar)     &(solar)      &              &                                   &(keV)&(solar)     &(solar)      &               \\
\hline
1&0.0-37.3   &$3.15^{+0.19}_{-0.21}$&$0.88^{+0.56}_{-0.45}$&$0.81^{+0.24}_{-0.20}$&43.9/42&&$2.80^{+0.40}_{-0.45}$&$1.25^{+1.80}_{-0.98}$&$1.25^{+0.92}_{-0.61}$&34.8/42\\
2&37.3-74.1  &$3.41^{+0.32}_{-0.22}$&$0.58^{+0.49}_{-0.43}$&$0.48^{+0.17}_{-0.15}$&40.6/43&&$3.14^{+0.44}_{-0.44}$&$0.67^{+1.05}_{-0.67}$&$0.61^{+0.40}_{-0.28}$&34.1/43\\
3&74.1-130.1 &$3.74^{+0.37}_{-0.37}$&$0.50^{+0.55}_{-0.46}$&$0.26^{+0.17}_{-0.14}$&35.5/42&&$3.28^{+0.69}_{-0.54}$&$0.69^{+0.93}_{-0.68}$&$0.18^{+0.29}_{-0.18}$&30.8/42\\ 
4&130.1-230.0&$4.39^{+0.54}_{-0.45}$&$ < 0.32 $            &$0.42^{+0.21}_{-0.20}$&37.2/37&&$4.37^{+0.70}_{-0.57}$&$ < 0.32$      &$0.52^{+0.31}_{-0.28}$&41.3/42 \\
5&230.0-444.1&$3.96^{+1.08}_{-0.78}$&$1.73^{+2.37}_{-1.00}$&$0.35^{+0.65}_{-0.31}$&30.1/41&&$-$                   & $-$                  &$-$                   &$-$     \\
\enddata
\tablenotetext{a}{Absorptions fixed to the Galactic value $N_{\rm H} = 4.04 \times 10^{20}$ cm$^{-2}$ (Dickey \& Lockman 1990).} 

\end{deluxetable}

\begin{deluxetable}{llcccccr}
\tablecaption{Fits to the Spectra Extracted in 0.0-74.1$h_{70}^{-1}$ kpc and 74.1-444.1$h_{70}^{-1}$ kpc}
\tabletypesize{\scriptsize}
\tablewidth{0pt}
\tablehead{
\colhead{Region} & 
\colhead{Radius} & 
\colhead{$N_{\rm{H}}$} & 
\colhead{$kT$} &
\colhead{$Z_{\alpha}$ \tablenotemark{a} \,} &
\colhead{$Z_{\rm{Si}}$} & 
\colhead{$Z_{\rm{Fe}}$} &
\colhead{$\chi^2_{\nu}$} \\ 
\colhead{} & 
\colhead{($h_{70}^{-1}$ kpc)} &
\colhead{$(10^{20}$ cm$^{-2})$} & 
\colhead{(keV)} & 
\colhead{(solar)} & 
\colhead{(solar)} &
\colhead{(solar)} & 
}
\startdata
Inner&0.0-74.1     &4.04 (Fixed)  &$3.20^{+0.15}_{-0.15}$&$0.87^{+0.52}_{-0.43}$&$0.48^{+0.37}_{-0.35}$&$0.62^{+0.14}_{-0.12}$&101.2/96\\
&          &$4.79^{+1.28}_{-1.24}$&$3.14^{+0.18}_{-0.19}$&$0.89^{+0.51}_{-0.42}$&$0.41^{+0.38}_{-0.34}$&$0.60^{+0.14}_{-0.12}$&100.2/95\\
Outer&74.1-444.1   &4.04 (Fixed)  &$3.81^{+0.32}_{-0.34}$&$1.17^{+0.73}_{-0.60}$&$0.37^{+0.44}_{-0.37}$&$0.31^{+0.14}_{-0.12}$&79.1/98\\
&          &$4.05^{+1.48}_{-1.45}$&$3.81^{+0.42}_{-0.44}$&$1.17^{+1.73}_{-0.60}$&$0.37^{+0.48}_{-0.37}$&$0.31^{+0.15}_{-0.13}$&79.1/97\\
\enddata
\tablenotetext{a}{Average abundance of the $\alpha$-elements except for Si.} 
\end{deluxetable}

\begin{deluxetable}{cccccccr}
\tablecaption{Fits to the Observed X-ray Surface Brightness Profile \label{tbl:gasmodel}}
\tabletypesize{\scriptsize}
\tablewidth{0pt}
\tablehead{
\multicolumn{4}{c}{$\beta$ Model} & &
\multicolumn{3}{c}{NFW Model} \\
\cline{1-4}
\cline{6-8}
$n_{g}$ (10$^{-3}$ cm$^{-3}$)&$R_{\rm c}$ ($h_{70}^{-1}$ kpc)&$\beta$&$\chi^2_{\nu}$&&$R_{\rm s}$ ($h_{70}^{-1}$ kpc)&$c$      &$\chi^2_{\nu}$\\
37.5$_{-1.2}^{+1.5}$&47.3$_{-0.3}^{+0.3}$  &0.612$^{+0.003}_{-0.003}$&101.1/62    &&195.1$_{-0.5}^{+0.5}$&$6.5_{-0.1}^{+0.1}$&69.2/62  \\
}
\startdata
\multicolumn{8}{c}{Two$-\beta$ Model} \\ \cline{1-8}
\hline
$n_{g,1}$ (10$^{-3}$ cm$^{-3}$)&$R_{\rm c1}$ ($h_{70}^{-1}$ kpc)&$\beta _{1}$&$n_{g,2}$ (10$^{-3}$ cm$^{-3}$)&&$R_{\rm c2}$ ($h_{70}^{-1}$ kpc)&$\beta_{2}$&$\chi^2_{\nu}$\\
21.7$_{-0.7}^{+0.8}$           &65.9$_{-0.5}^{+0.6}$          &0.660$^{+0.004}_{-0.005}$&40.4$_{-1.2}^{+1.5}$&&15.5$_{-0.4}^{+0.3}$            &0.586$^{+0.005}_{-0.011}$& 46.1/59\\
\enddata
\end{deluxetable}

\begin{deluxetable}{lccccr}
\tablecaption{Iron Enrichment by SNe Ia \label{tbl:enrichtime}}
\tabletypesize{\scriptsize}
\tablewidth{0pt}
\tablehead{
\colhead{SN Ia Rate} &
\multicolumn{2}{c}{$t_{\rm enrich}(10^{9}~{\rm yr})$} & &
\multicolumn{2}{c}{$z$ \tablenotemark{a} \,} \\ 
\cline{2-3}
\cline{5-6}
\colhead{($h^{2}_{70}$ $\rm SNu$)} & 
\colhead{$\eta_{\rm Fe}=0.5$ $\rm M_{\odot}$} &
\colhead{$0.7$ $\rm M_{\odot}$} & &
\colhead{$\eta_{\rm Fe}=0.5$ $\rm M_{\odot}$} &
\colhead{$0.7$ $\rm M_{\odot}$}
}
\startdata
      0.18 
      &$-$&$9.21^{+2.36}_{-1.94}$&&$-$&$> 1.66$\\
\hline
      Observed \tablenotemark{b} \,
      &$> 6.78$&$6.36^{+0.99}_{-0.87}$&&$> 1.45$&$1.30^{+0.41}_{-0.26}$\\
\enddata
\tablenotetext{a}{Redshift at which SN Ia iron enrichment is expected to start.}
\tablenotetext{b}{Data are from Dahlen et al. (2004).}
\end{deluxetable}

\clearpage

{\bf Figure Captions}

\noindent Fig.1 -- 
(a): Central $4.9^{\prime}$ ($944h_{70}^{-1}$ kpc) of MS 0839.9+2938 in 0.7--8 keV, which is 
plotted in the logarithmic scale. The image has been background-subtracted, exposure-corrected, 
and smoothed with a minimum significance of 3 and a maximum significance of 5. X-ray intensity 
contours are spaced from 
$1.4\times10^{-10}$ 
to 
$4.7\times10^{-7}$ photons cm$^{-2}$ s$^{-1}$ pixel$^{-2}$. 
The inner $1^{\prime}$ region, which is marked by a box, is magnified in (b), where the contour 
levels are from 
$5.7\times10^{-9}$ 
to 
$2.5\times10^{-7}$ photons cm$^{-2}$ s$^{-1}$ pixel$^{-2}$. 
(c): Corresponding DSS optical image for the same sky field as in (a) on which the X-ray contours 
are overlaid for comparison.

\noindent Fig.2 -- Azimuthally-averaged radial surface brightness profile of MS 0839.9+2938 in 
0.7--8 keV. The background is assumed to be spatially uniform and is represented by the dotted 
line. In $15.8-800h_{70}^{-1}$ kpc, the surface brightness profile can be well described by the 
$\beta$ model (dashed line), showing an obvious central emission excess. The fits of the two-$\beta$, 
$\beta$ and NFW models to the whole cluster region, and the fit of the $\beta$ model to the outer
region only are shown with the solid, dashed, dot-dashed and dot-dot-dot-dashed lines, respectively.

\noindent Fig.3 -- Two-dimensional distribution of the hardness-ratio of the diffuse emission 
in MS 0839.9+2938, which is defined as the ratio of the background-subtracted and exposure-corrected 
counts in 1.4--8.0 keV to those in 0.7--1.4 keV. The cross represents the peak of X-ray emission, 
and the circle ($r=74h_{70}^{-1}$ kpc) marks the region in which significant temperature drop and 
abundance increase are detected. The image is plotted in linear scale, and has been smoothed in 
the same way as in Figure 1.

\noindent Fig.4 -- (a): Projected (solid) and deprojected (dashed) radial distributions of the 
gas temperature, which are obtained with the absorbed APEC model. The best-fit temperature profile 
for the deprojected distribution is shown as a dotted line. 
(b): The same as (a) but for the absorbed VAPEC model. 
(c): Projected (solid) and deprojected (dashed) radial distributions of the average metal 
abundance, which are obtained with the absorbed APEC model. 
(d): The same as (c) but for the iron abundance obtained with the absorbed VAPEC model. 

\noindent Fig.5 -- Confidence contours at the 68.3\%, 90\% and 99\% confidence levels for 
the gas temperature and iron abundance derived with the deprojected VAPEC model for the 
innermost ($R < 37.3h_{70}^{-1}$ kpc) and the third ($74.1 < R < 130.1h_{70}^{-1}$ kpc)
annular regions. 

\noindent Fig.6 -- Upper and middle panels: calculated radial distributions of the total 
gravitating mass, gas mass and gas mass fraction, along with the 90\% error limits (shown 
as dotted lines). The total mass was calculated by using both $\beta$ and the best-fit 
two-$\beta$ model. The gas mass and gas fraction were calculated by using the best-fit 
two-$\beta$ model. Lower panel: the calculated cooling time as a function of radius. The 
horizontal, dashed line represents the age of the universe at the redshift of cluster. 
Radii within which the gas temperature drops and iron abundance increases significantly
($R_{\rm a}$), and at which the gas temperature begins to decrease inwards ($R_{\rm b}$), 
are indicated by vertical lines.

\noindent Fig.7 -- Upper and middle panels: Spectrum extracted in $<74.1 h_{70}^{-1}$ kpc, 
where the gas temperature is significantly lower than that of the outer regions, along 
with the best-fit two-temperature APEC model and residuals. Lower panel: model ratios 
of the single MKCFLOW model and MEKAL+MKCFLOW model to the two-temperature APEC model. 
$kT_{\rm low}$ of the MKCFLOW component is fixed to 0.08 keV.

\noindent Fig.8 -- Ratios of the iron abundance in cluster's inner region to that in the outer
region Z$_{\rm in}$/Z$_{\rm out}$ as a function of the cluster's (a) average gas temperature and (b) 
optical luminosity for nearby clusters (solid; Tamura et al 2004; Girardi et al. 2000; 
David et al. 1995; Pratt \& Arnaud 2003) and two intermediately distant clusters, i.e.,  
MS 0839.9+2938 (dashed) and Abell 2390 (dot-dashed. This is the average abundance ratio rather 
than iron abundance ratio; Allen et al. 2001). The ratio for the nearby clusters 
is shown as a horizontal dotted line.

\noindent Fig.9 -- Radial distribution of the gas entropy (crosses and the dashed line) along 
with the theoretical expectation quoted from Tozzi et al. 2001 (solid line).

\clearpage

\begin{figure}
\epsscale{1.0}
\begin{center}
\includegraphics[width=5.0cm,angle=0]{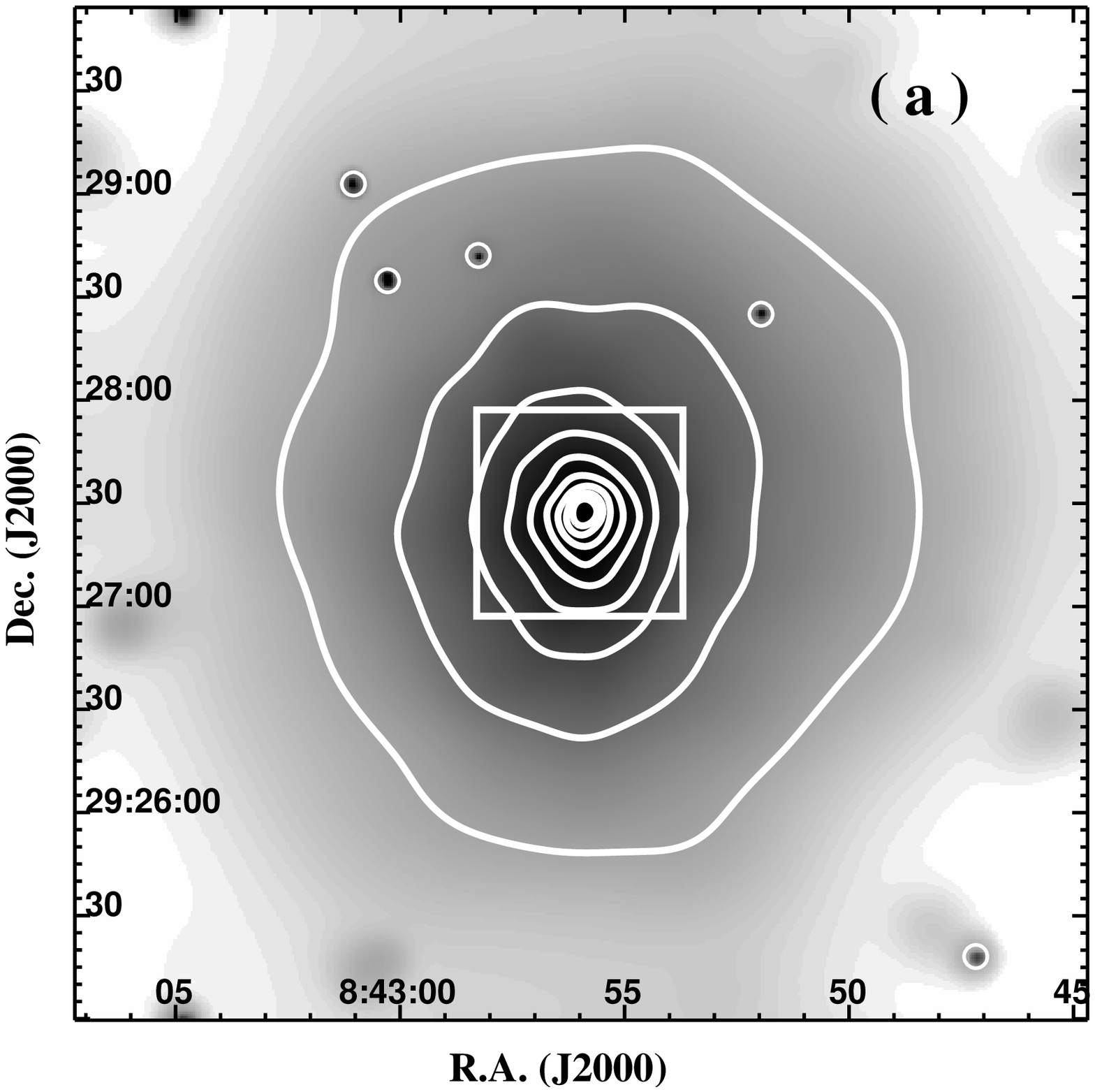}
\includegraphics[width=5.0cm,angle=0]{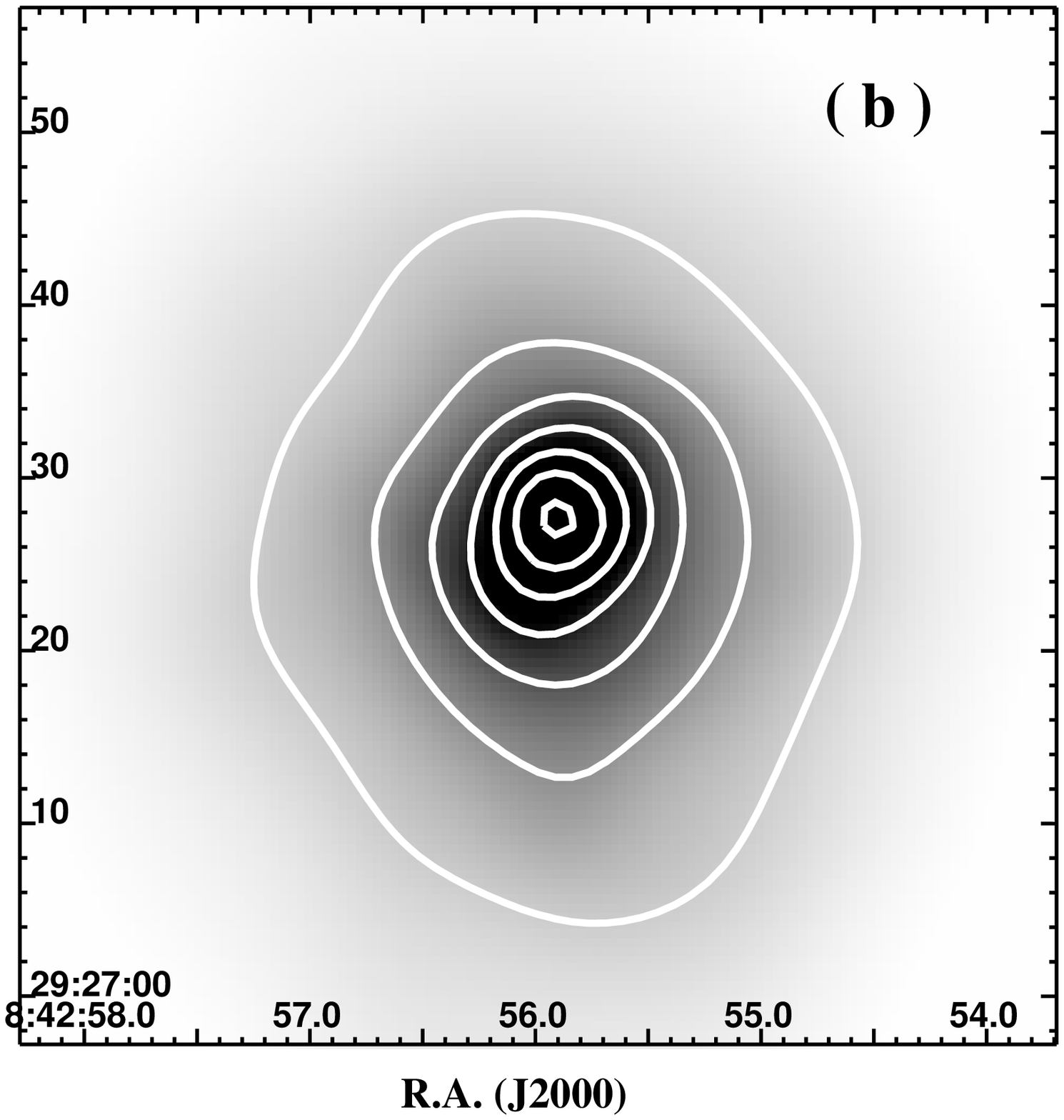}
\includegraphics[width=5.0cm,angle=0]{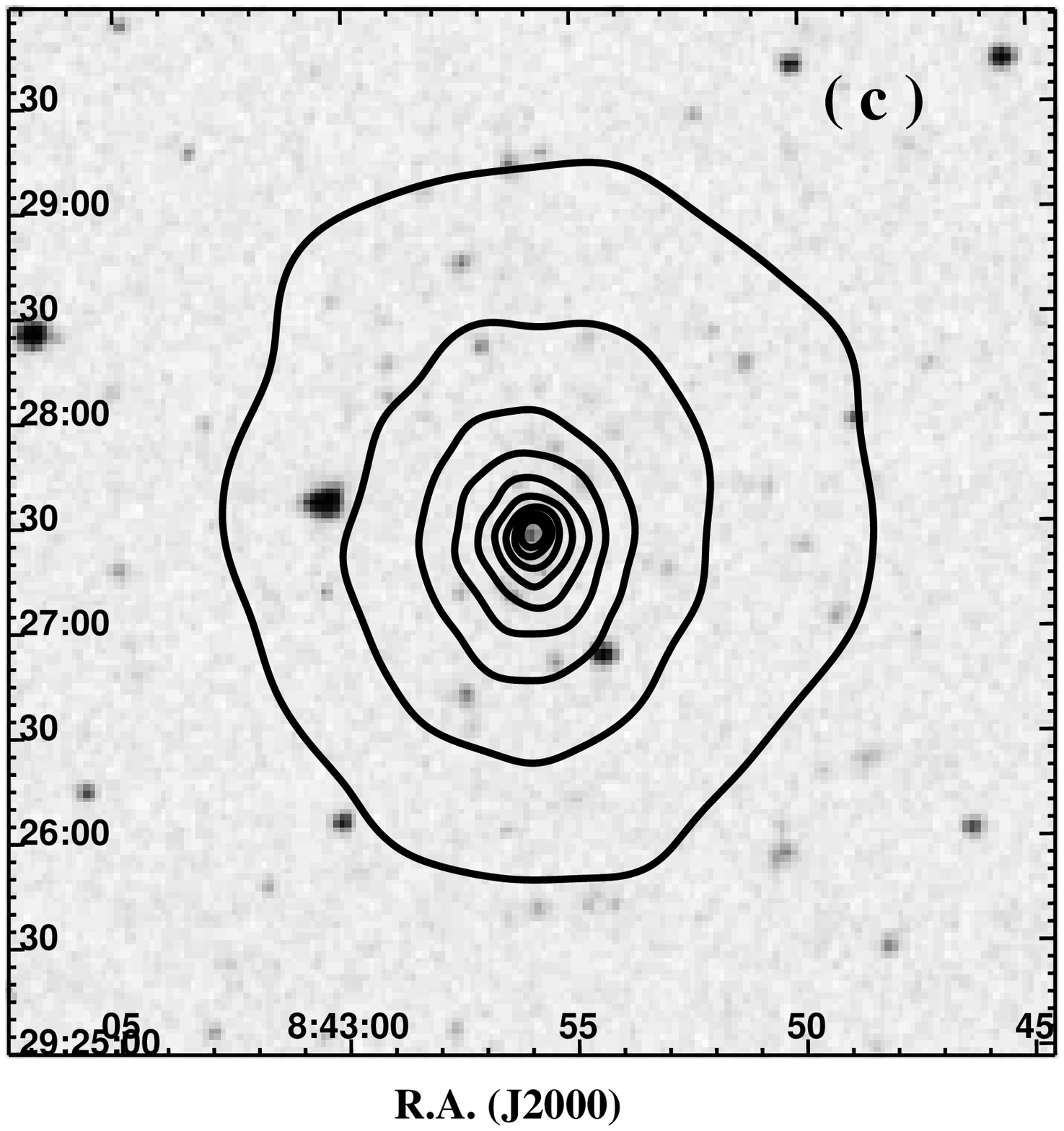}
\end{center}
\figcaption{
(a): Central $4.9^{\prime}$ ($944h_{70}^{-1}$ kpc) of MS 0839.9+2938 in 0.7--8 keV, which is 
plotted in the logarithmic scale. The image has been background-subtracted, exposure-corrected, 
and smoothed with a minimum significance of 3 and a maximum significance of 5. X-ray intensity 
contours are spaced from 
$1.4\times10^{-10}$ 
to 
$4.7\times10^{-7}$ photons cm$^{-2}$ s$^{-1}$ pixel$^{-2}$. 
The inner $1^{\prime}$ region, which is marked by a box, is magnified in (b), where the contour 
levels are from 
$5.7\times10^{-9}$ 
to 
$2.5\times10^{-7}$ photons cm$^{-2}$ s$^{-1}$ pixel$^{-2}$. 
(c): Corresponding DSS optical image for the same sky field as in (a) on which the X-ray contours 
are overlaid for comparison.
\label{fig1}}
\end{figure}

\begin{figure}
\begin{center}
\includegraphics[width=5.5cm,angle=270]{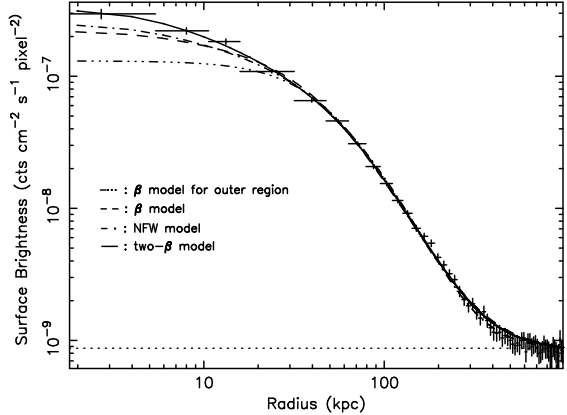}
\end{center}
\figcaption{
Azimuthally-averaged radial surface brightness profile of MS 0839.9+2938 in 
0.7--8 keV. The background is assumed to be spatially uniform and is represented by the dotted 
line. In $15.8-800h_{70}^{-1}$ kpc, the surface brightness profile can be well described by the 
$\beta$ model (dashed line), showing an obvious central emission excess. The fits of the two-$\beta$, 
$\beta$ and NFW models to the whole cluster region, and the fit of the $\beta$ model to the outer
region only are shown with the solid, dashed, dot-dashed and dot-dot-dot-dashed lines, respectively.
\label{fig2}}
\end{figure}

\begin{figure}
\begin{center}
\includegraphics[width=7.cm,angle=0]{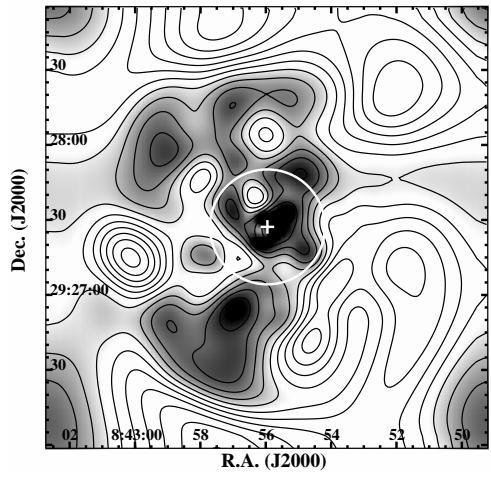}
\end{center}
\figcaption{
Two-dimensional distribution of the hardness-ratio of the diffuse emission 
in MS 0839.9+2938, which is defined as the ratio of the background-subtracted and exposure-corrected 
counts in 1.4--8.0 keV to those in 0.7--1.4 keV. The cross represents the peak of X-ray emission, 
and the circle ($r=74h_{70}^{-1}$ kpc) marks the region in which significant temperature drop and 
abundance increase are detected. The image is plotted in linear scale, and has been smoothed in 
the same way as in Figure 1.
\label{fig3}}
\end{figure}

\begin{figure}
\begin{center}
\includegraphics[width=7.0cm,angle=270]{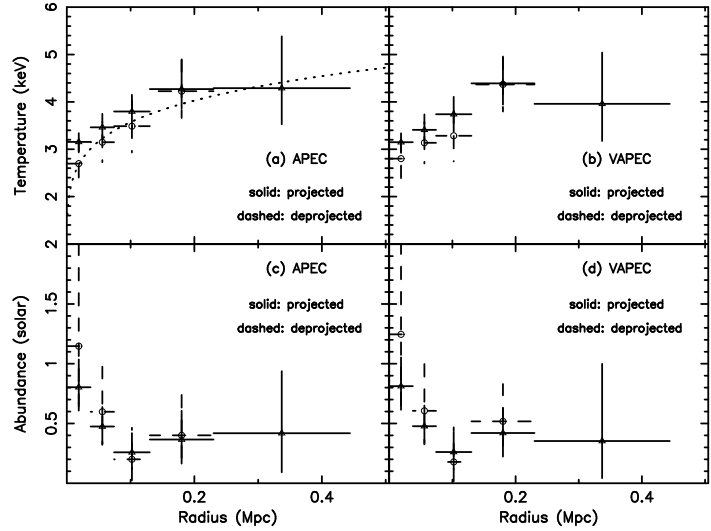}
\end{center}
\figcaption{
(a): Projected (solid) and deprojected (dashed) radial distributions of the 
gas temperature, which are obtained with the absorbed APEC model. The best-fit temperature profile 
for the deprojected distribution is shown as a dotted line. 
(b): The same as (a) but for the absorbed VAPEC model. 
(c): Projected (solid) and deprojected (dashed) radial distributions of the average metal 
abundance, which are obtained with the absorbed APEC model. 
(d): The same as (c) but for the iron abundance obtained with the absorbed VAPEC model. 
\label{fig4}}
\end{figure}

\begin{figure}
\epsscale{0.6}
\begin{center}
\includegraphics[width=6.0cm,angle=270]{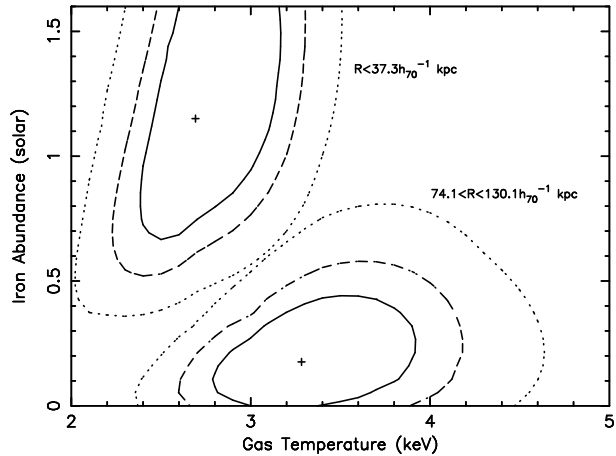}
\end{center}
\figcaption{
Confidence contours at the 68.3\%, 90\% and 99\% confidence levels for 
the gas temperature and iron abundance derived with the deprojected VAPEC model for the 
innermost ($R < 37.3h_{70}^{-1}$ kpc) and the third ($74.1 < R < 130.1h_{70}^{-1}$ kpc)
annular regions. 
\label{fig5}}
\end{figure}

\begin{figure}
\epsscale{0.6}
\begin{center}
\includegraphics[width=5.5cm,angle=270]{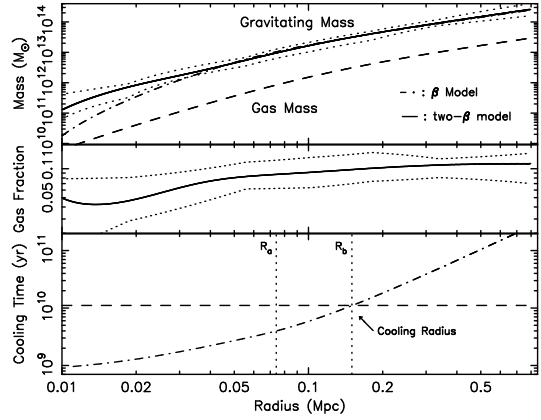}
\end{center}
\figcaption{
Upper and middle panels: calculated radial distributions of the total 
gravitating mass, gas mass and gas mass fraction, along with the 90\% error limits (shown 
as dotted lines). The total mass was calculated by using both $\beta$ and the best-fit 
two-$\beta$ model. The gas mass and gas fraction were calculated by using the best-fit 
two-$\beta$ model. Lower panel: the calculated cooling time as a function of radius. The 
horizontal, dashed line represents the age of the universe at the redshift of cluster. 
Radii within which the gas temperature drops and iron abundance increases significantly
($R_{\rm a}$), and at which the gas temperature begins to decrease inwards ($R_{\rm b}$), 
are indicated by vertical lines.
\label{fig6}}
\end{figure}

\begin{figure}
\epsscale{0.6}
\begin{center}
\includegraphics[width=5.5cm,angle=270]{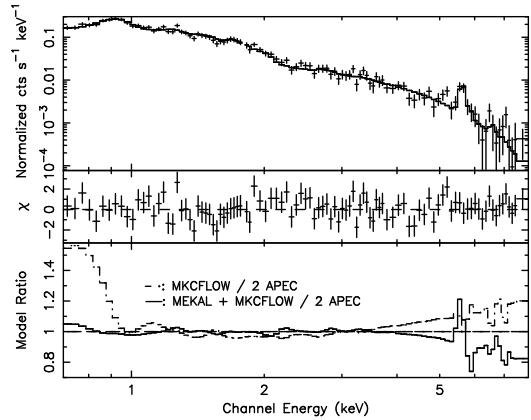}
\end{center}
\figcaption{
Upper and middle panels: Spectrum extracted in $<74.1 h_{70}^{-1}$ kpc, 
where the gas temperature is significantly lower than that of the outer regions, along 
with the best-fit two-temperature APEC model and residuals. Lower panel: model ratios of 
the single MKCFLOW model and MEKAL+MKCFLOW model to the two-temperature APEC model. 
$kT_{\rm low}$ of the MKCFLOW component is fixed to 0.08 keV.
\label{fig7}}
\end{figure}

\begin{figure}
\begin{center}
\includegraphics[width=5.5cm,angle=270]{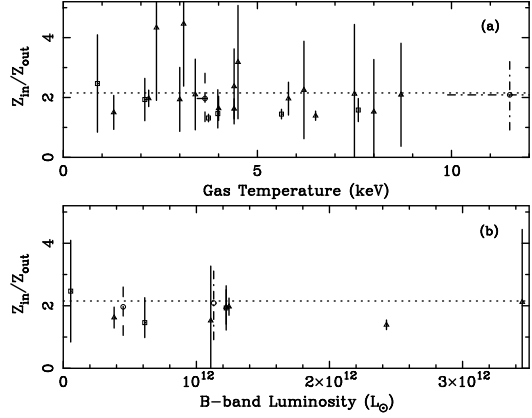}
\end{center}
\figcaption{
Ratios of the iron abundance in cluster's inner region to that in the outer
region Z$_{\rm in}$/Z$_{\rm out}$ as a function of the cluster's (a) average gas temperature and (b) 
optical luminosity for nearby clusters (solid; Tamura et al 2004; Girardi et al. 2000; 
David et al. 1995; Pratt \& Arnaud 2003) and two intermediately distant clusters, i.e.,  
MS 0839.9+2938 (dashed) and Abell 2390 (dot-dashed. This is the average abundance ratio rather than 
iron abundance ratio; Allen et al. 2001). The mean ratio for the nearby clusters is shown 
as a horizontal dotted line.
\label{fig8}}
\end{figure}

\begin{figure}
\epsscale{0.6}
\begin{center}
\includegraphics[width=5.5cm,angle=270]{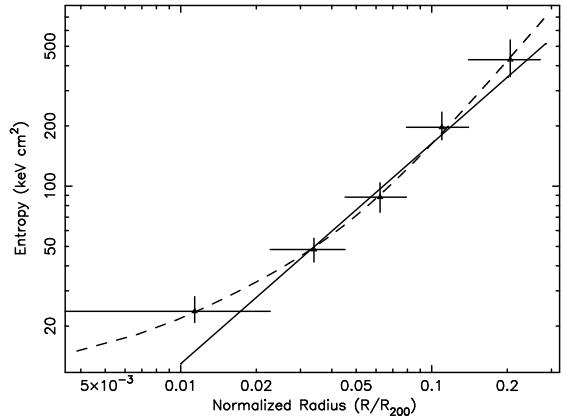}
\end{center}
\figcaption{
Radial distribution of the gas entropy (crosses and the dashed line) along 
with the theoretical expectation quoted from Tozzi et al. 2001 (solid line).
\label{fig9}}
\end{figure}

\end{document}